\documentclass[preprint,showpacs]{revtex4}

\usepackage{graphicx}
\usepackage{bm}

\usepackage{amsmath}

\begin{document}


\title{
Spin-dependent three-nucleon force effects on nucleon-deuteron scattering
}

\author{S. Ishikawa}\email[E-mail:]{ishikawa@hosei.ac.jp}
\affiliation{
Department of Physics, Science Research Center, Hosei University, 
2-17-1 Fujimi, Chiyoda, Tokyo 102-8160, Japan} 
\affiliation{
Triangle Universities Nuclear Laboratory, 
Durham, North Carolina 27708-0308, U.S.A.} 

\date{\today}

\begin{abstract} 
We construct a phenomenological three-nucleon force (3NF) model that gives a good description of polarization observables in elastic nucleon-deuteron ($N$-$d$) scattering at a low energy together with a realistic nucleon-nucleon force and a 3NF arising from the exchange of two pions.
Parameters of the model, which consists of spin-independent, spin-orbit, and tensor components, are determined to reproduce the three-nucleon binding energy and polarization observables in $N$-$d$ scattering at 3 MeV. 
Predictions of the model 3NF on $N$-$d$ polarization observables at higher energies are examined, and effects of each component on the observables are investigated.
\end{abstract}

%





%

\pacs{21.30.-x, 21.45.+v, 24.70.+s}

\maketitle


As is well known, modern two-nucleon force (2NF) models have a deficiency in explaining the binding energies of three-nucleon (3N) systems, and this problem is successfully solved by introducing a 3NF arising from the exchange process of two pions among three nucleons, which is called the two pion-exchange (2$\pi$E) 3NF \cite{Co79,Co83}.
However, such combinations of the 2NFs and the 2$\pi$E-3NF that reproduce the 3N binding energy do not necessarily explain polarization observables in 3N scattering systems such as vector or tensor analyzing powers in elastic $N$-$d$ scattering. 
See, e.g., Table III of Ref. \cite{Ki01}, where calculations of observable with and without a 2$\pi$E-3NF are compared in terms of $\chi^2$ to experimental data below 30 MeV of incident nucleon energy in laboratory system. 
In spite of recent progress in constructing realistic 3NFs from chiral effective field theory or from heavier-boson-exchange mechanisms, no consensus has been obtained for possible mechanisms of 3NFs consistent with all of the experimental data. 
On the other hand, model 3NFs with artificial functional forms have been proposed to explain the polarization observables quite well \cite{Ki99,Is03,Is04}. 
These 3NFs have a form that typical components in 2NFs, e.g., central spin-independent, tensor, or spin-orbit components, are modified in the presence of third nucleon. (See Eq. (\ref{eq:V_3nf_ph}) below.) 
In this paper, we introduce such a phenomenological 3NF to resolve the discrepancies by a 2NF and the 2$\pi$E-3NF at a low energy, and examine whether it is still valid or not for $N$-$d$ observables at higher energies up to 30 MeV. 
Since it may not be so difficult to understand what physical process is simulated by the spin dependence of each component, we expect that the present study will provide some hints for characteristics about more realistic 3NFs to be studied.

Our 3N calculations are based on a formalism to solve the Faddeev equations in coordinate space as integral equations \cite{Is87,Sa86}.
For scattering states below the 3N breakup threshold energy, effects of the long-range Coulomb force between two protons are exactly treated \cite{Is03d}. 
Calculations for energies above the 3N breakup threshold are formulated in Ref. \cite{Is07}. 
3N partial wave states for which 2NFs and 3NFs act, are restricted to those with total two-nucleon angular momenta $j \le 6$ for bound state calculations, and $j \le 3$ for scattering state calculations.
The total 3N angular momentum ($J$) is truncated at $J=19/2$, while 3NFs are switched off for 3N states with $J>9/2$. 
These truncating procedures are confirmed to give converged results for the aim of the present work. 

We use the Argonne V$_{18}$ model (AV18) \cite{Wi95} for the realistic 2NF and the Brazil model (BR) \cite{Co83} for the 2$\pi$E-3NF. 
Calculated triton binding energies with models used in this work are tabulated in Table \ref{tab:b3}. 
The AV18 calculation underbinds the triton by about 0.9 MeV. 
The introduction of the 2$\pi$E-3NF produces enough attraction to remedy the defect, but it strongly depends on the choice of cutoff parameter in the $\pi NN$ form factor. 
We choose a dipole form factor, $\left(\frac{\Lambda^2-m_\pi^2}{\bm{q}^2+\Lambda^2}\right)^2$, where $\bm{q}$ is the momentum of the exchanged pion, $m_\pi$ the pion mass, and $\Lambda$ the cutoff mass.
The choice of $\Lambda = 800$ MeV $\approx 5.8 m_\pi$ (BR$_{800}$), which is  close to a value cited in Ref. \cite{Co81,Co83,Jo75} to explain the Goldberger-Treiman discrepancy,  overshoots the triton binding energy by about 0.9 MeV. 
It turns out that the binding energy is reproduced when we take $\Lambda = 680$ MeV $\approx 4.9 m_\pi$ (BR$_{680}$). 
This rather small value of $\Lambda$ may be considered as a result of incorporating unknown 3NF effects.

\begin{table}[b]
\caption{\label{tab:b3}
Empirical value and calculated values of the triton binding energy ($B_3$).
See the text for the description of the models.
}
\begin{tabular}{lc}
\hline\hline
Model  &  $B_3$ (MeV) \\
\hline
Empirical & 8.482 \\
AV18      & 7.626 \\
AV18+BR${}_{800}$       &  9.380 \\
AV18+BR${}_{680}$       & 8.493 \\
AV18+BR$_{800}$+C+T+SO  & 8.478 \\
AV18+BR${}_{680}$+SO    & 8.444 \\
\hline\hline
\end{tabular}
\end{table}


\begin{figure}[tb]
\includegraphics[scale=0.28]{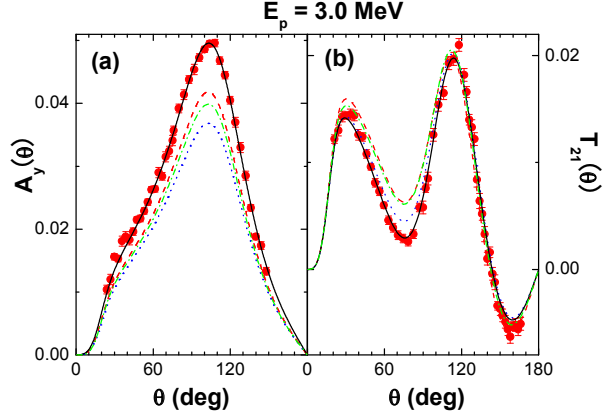}
\caption{(Color online) 
(a) Proton vector analyzing power $A_y(\theta)$ and (b) deuteron tensor analyzing power $T_{21}(\theta)$ in $p$-$d$ scattering at $E_p=3.0$ MeV ($E_d=6.0$ MeV). 
Dotted (blue) curves denote the AV18 calculations;
dashed (red) curves AV18+BR$_{800}$; 
dotted-dashed (green) curves AV18+BR$_{680}$; 
solid (black) curves AV18+BR$_{800}$+C+T+SO.
Solid circles are experimental data from Ref. \protect\cite{Sa94,Sh95}.
\label{fig:ay-t21-3mev}
}
\end{figure}

As examples to demonstrate effects of 3NFs at low energy, we show our calculations for the proton vector analyzing power $A_y(\theta)$ and the deuteron tensor analyzing power $T_{21}(\theta)$ in elastic proton-deuteron ($p$-$d$) scattering at $E_p=3.0$ MeV (or $E_d=6.0$ MeV) in Fig. \ref{fig:ay-t21-3mev} comparing with experimental data \cite{Sa94,Sh95}. 
Deficiencies of the AV18 calculations in reproducing the data prominently appear as a smallness for the maximum of $A_y(\theta)$ at $\theta\approx 100^\circ$ and as an excess for the local minimum of $T_{21}(\theta)$ at $\theta\approx 80^\circ$. 
The AV18+BR$_{800}$ and the AV18+BR$_{680}$ calculations, which almost agree, demonstrate that the 2$\pi$E-3NF partially remedies the deficiency of the $A_y$-maximum but worsens that of the $T_{21}$-minimum.
In Ref. \cite{Is03}, the latter effect was shown to arise from a tensor component in the 2$\pi$E-3NF.

Now we consider a phenomenological 3NF to improve the results of AV18+BR$_{800}$ in Fig. \ref{fig:ay-t21-3mev} in the following form \cite {Ki99,Is03,Is04}, 
\begin{eqnarray}
V &=& \sum_{i < j} 
  e^{-(\frac{r_{ik}}{r_G})^2 -(\frac{r_{jk}}{r_G})^2}
 \left\{ V_0 + V_T S_T(ij) {\hat P}_{11}(ij) \right\}
\nonumber \\
 &&+  V_{ls} e^{-\alpha\rho} 
  \sum_{i < j} [\bm{l}_{ij} \cdot (\bm{S}_i+\bm{S}_j)] 
   \hat{P}_{11}(ij),
\label{eq:V_3nf_ph}
\end{eqnarray}
where $S_T(ij)$ is the tensor operator acting between nucleon pair $(i,j)$, $\hat{P}_{11}(ij)$ the projection operator to the spin and isospin triplet state of the $(i,j)$ pair, and $\rho^2=\frac23(r_{12}^2+r_{23}^2+r_{31}^2)$. 
In the present work, the range parameter $r_G$ is taken to be 1.0 fm as in Refs. \cite{Is03,Is04}, and $\alpha$ to be 1.5 fm${}^{-1}$, which is the shortest one in Ref. \cite{Ki99}. 
The parameters for the strength of the central spin-independent component (C) $V_0$, the tensor component (T) $V_T$, and the spin-orbit component (SO) $V_{ls}$ are determined to reproduce the triton binding energy and the observables in Fig. \ref{fig:ay-t21-3mev} as follows: 
In Ref. \cite{Is04}, it is reported that one can simulate the 2$\pi$E-3NF by Eq. (\ref{eq:V_3nf_ph}) with a choice of $(V_0, V_T) = (-38~\text{MeV}, +20~\text{MeV})$, and then get a remarkable improvement in $T_{21}(\theta)$ by changing the sign of $V_T$ and readjusting $V_0$ to fit the triton binding energy.
Thus, in the present case, we choose the variation $-20-(+20)=-40$ MeV for the value of $V_T$ to improve $T_{21}(\theta)$ in addition to AV18+BR$_{800}$. 
Hereafter, we call this procedure as {\em tensor inversion}.
The strength of the spin-orbit component $V_{ls}$ is determined to be -16 MeV  to reproduce $A_y(\theta)$ at 3.0 MeV.
Finally, the strength of the spin-independent component $V_0$ is determined to reproduce the triton binding energy. 
The values obtained are $(V_0, V_T, V_{ls}) = (+25~\text{MeV}, -40~\text{MeV}, -16~\text{MeV})$. 
The results of this 3NF (AV18+BR$_{800}$+C+T+SO) are displayed as the solid curves in Fig. \ref{fig:ay-t21-3mev}.

We give a remark here that the C-3NF plays an essential role in reproducing the binding energy to get the repulsive effect against the large attraction from the BR$_{800}$-3NF. 
To the total repulsion of about 0.90 MeV by the C+T+SO-3NF, (see Table \ref{tab:b3},) the C-3NF contributes about 0.62 MeV, the T-3NF does about $0.24$ MeV, and the SO-3NF does 0.05 MeV.

In order to understand a role of the tensor inversion, we make another model 3NF so that $A_y(\theta)$ at 3.0 MeV is reproduced with AV18+BR$_{680}$ plus Eq. (\ref{eq:V_3nf_ph}) with $V_0=V_T=0$, which will be denoted by AV18+BR$_{680}$+SO. 
In this case, the strength of the SO term becomes $V_{ls} = -21$ MeV. 
Since the effect of the SO-3NF on the triton binding energy is rather small, about 0.05 MeV of repulsive contribution, AV18+BR$_{680}$+SO reproduces the binding energy as well.
In Fig.  \ref{fig:ay-t21-3mev}, the results of AV18+BR$_{680}$+SO are not plotted because they coincide with the AV18+BR$_{800}$+C+T+SO calculation for $A_y(\theta)$, and with AV18+BR$_{680}$ for $T_{21}(\theta)$.
Note that the latter demonstrates that the SO-3NF gives only minor effect on $T_{21}(\theta)$.


\begin{figure}[tb]
\includegraphics[scale=0.28]{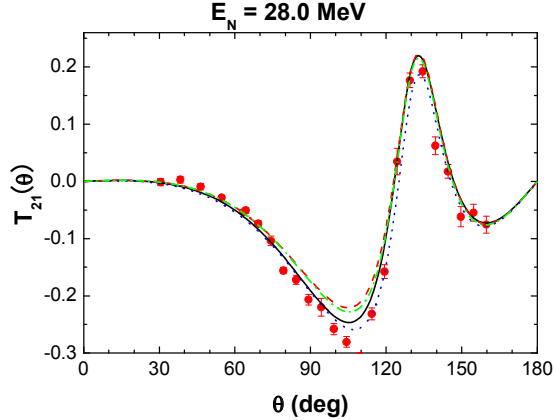}
\caption{(Color online) 
Deuteron tensor analyzing power $T_{21}(\theta)$ in $N$-$d$ scattering at $E_N=28.0$ MeV (or $E_d=56.0$ MeV). 
Calculations are for $n$-$d$: 
dotted (blue) curve denotes the AV18 calculation; 
dashed (red) curve AV18+BR$_{680}$+SO; 
dotted-dashed (green) curve AV18+BR$_{680}$; 
solid (black) curve AV18+BR$_{800}$+C+T+SO.
Solid circles are experimental data for $p$-$d$ scattering from Ref. \protect\cite{Ha84}.
\label{fig:t21-28mev}
}
\end{figure}

%
In Fig. \ref{fig:t21-28mev}, we show calculations of the tensor analyzing power $T_{21}(\theta)$ in elastic neutron-deuteron ($n$-$d$) scattering at $E_n = 28.0$ MeV (or $E_d=56.0$ MeV) comparing with $p$-$d$ data at the corresponding energy \cite{Ha84}, for which the effect of the Coulomb force may be small as shown in Ref. \cite{Ki01}. 
In this energy, the AV18 calculation looks to be almost in agreement with the data. 
However, the introduction of the 2$\pi$E-3NF destroys the fit as demonstrated by the dotted-dashed curve. 
The small difference between AV18+BR$_{680}$ and AV18+BR$_{680}$+SO shows that the SO-3NF plays only a minor role in this observable as well as for 3.0 MeV.
On the other hand, the tensor inversion effect almost cancels the unfavorable effect due to the 2$\pi$E-3NF. 
It is noted that the effect of the tensor inversion in this energy is different from that in 3.0 MeV, which overshoots the effect of 2$\pi$E-3NF, although both effects are favorable in explaining the data.
This difference may occur because of a partial cancellation between the effect of the C-3NF and that of the T-3NF at higher energies.


\begin{figure}[tb]
\includegraphics[scale=0.29]{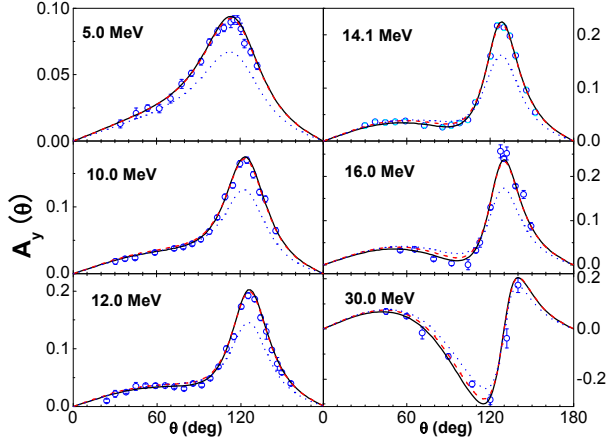}
\caption{(Color online) 
Neutron vector analyzing power $A_{y}(\theta)$ in $n$-$d$ scattering. 
Dotted (blue) curves denotes the AV18 calculations; 
dashed (red) curves AV18+BR$_{680}$+SO; 
solid (black) curves  AV18+BR$_{800}$+C+T+SO.
Experimental data are from Ref. \protect\cite{To91} for 5.0 MeV; 
from Ref. \protect\cite{To82} for 10.0 MeV; 
from Ref. \protect\cite{To83} for 14.1 MeV; 
from Ref. \protect\cite{Fu99} for 16.0 MeV; 
and from Ref. \protect\cite{Do78} for 30.0 MeV.
\label{fig:ay-e-dep}
}
\end{figure}

In Fig. \ref{fig:ay-e-dep}, we plot results of the neutron vector analyzing power $A_y(\theta)$ in $n$-$d$ scattering comparing with available experimental data at several energies up to 30 MeV \cite{To91,To82,To83,Fu99,Do78}.
First, we take an overview of the energy dependence in the $A_y(\theta)$ angular distribution. 
For lower energies,  $A_y(\theta)$ has a single peak at $\theta \approx 100^\circ$. 
As the energy increases, the angle where $A_y(\theta)$ takes the maximum, $\theta_{max}$, increases gradually up to about $140^\circ$ at $E_n=30$ MeV, and a plateau region appears at $\theta \approx 90^\circ$ for $E_n \geq 10$ MeV developing into a local minimum. 

Failures of the AV18 calculations in reproducing the minimum and the maximum of the $A_y(\theta)$ data are well recovered by the AV18+BR$_{800}$+C+T+SO and the AV18+BR$_{680}$+SO calculations. 
This is essentially because of the contribution from the SO-3NF.

\begin{figure}[tb]
\includegraphics[scale=0.29]{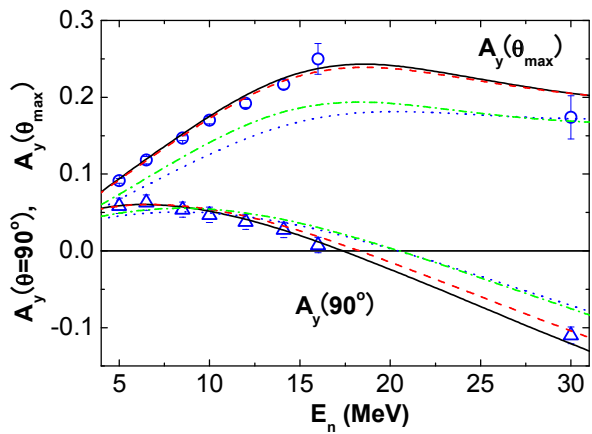}
\caption{(Color online) 
Neutron vector analyzing power $A_{y}(\theta)$ at $\theta=90^\circ$ and $\theta=\theta_{max}$ in $n$-$d$ scattering. 
Dotted (blue) curves denotes the AV18 calculations;
dotted-dashed (green) curves AV18+BR$_{680}$; 
dashed (red) curves AV18+BR$_{680}$+SO; 
solid (black) curves  AV18+BR$_{800}$+C+T+SO.
Points are calculated from experimental data \protect\cite{To91,To82,To83,Fu99,Do78}.
\label{fig:ay-90-max}
}
\end{figure}

A closer look at Fig. \ref{fig:ay-e-dep} shows that the AV18+BR$_{800}$+C+T+SO and the AV18+BR$_{680}$+SO calculations equivalently reproduce the $A_y$-maximum, but they display a difference at higher energies for angles $80^\circ \le \theta \le 120^\circ$, where $A_y(\theta)$ has the local minimum. 
In order to emphasize this, in Fig. \ref{fig:ay-90-max}, we plot the values of $A_y(\theta)$ at $\theta=90^\circ$ and those at $\theta_{max}$ as a function of the incident neutron energy $E_n$ comparing with values extracted from the experimental data \cite{To91,To82,To83,Fu99,Do78}. 
This figure clearly shows that the effect of the SO-3NF, which is displayed by the difference between the AV18+BR$_{680}$ and the AV18+BR$_{680}$+SO calculations, tends to improve $A_y(\theta)$ dominantly at both angles.
On the other hand, the tensor inversion effect, displayed by the difference between AV18+BR$_{680}$+SO and AV18+BR$_{800}$+C+T+SO, is small at $\theta_{max}$, but gives a nonnegligible contribution at $\theta=90^\circ$ as the energy increases, which works in the same direction as the SO-3NF.


\begin{figure}[tb]
\includegraphics[scale=0.29]{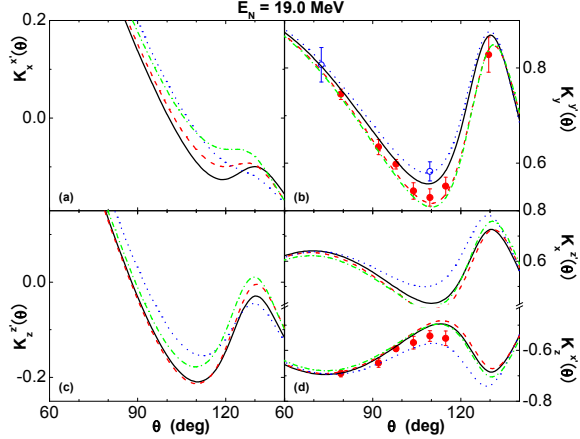}
\caption{(Color online) 
Nucleon to nucleon polarization transfer coefficients (a) $K_x^{x^\prime}(\theta)$, (b) $K_y^{y^\prime}(\theta)$, (c) $K_z^{z^\prime}(\theta)$,  and (d) $K_x^{z^\prime}(\theta)$ (upper) and $K_z^{x^\prime}(\theta)$ (lower) in $N$-$d$ scattering at $E_N = 19.0$ MeV.
Calculations are for $n$-$d$: 
dotted (blue) curves denotes the AV18 calculations; 
dotted-dashed (green) curves AV18+BR$_{680}$; 
dashed (red) curves AV18+BR$_{680}$+SO;
solid (black) curves AV18+BR$_{800}$+C+T+SO.
Open circles in (b) are experimental data for $n$-$d$ $K_y^{y^\prime}(\theta)$ \protect\cite{He98}; 
solid circles in (b) $p$-$d$ $K_y^{y^\prime}(\theta)$ \protect\cite{Sy94}; 
solid circles in (d) $p$-$d$ $K_z^{x^\prime}(\theta)$ \protect\cite{Sy94}. 
\label{fig:spin-trns-prime-19mev}
}
\end{figure}

As an another interesting example, we pickup nucleon-to-nucleon polarization transfer coefficients in $\vec{N}+d \to \vec{N}+d$ reaction, $K_x^{x^\prime}(\theta)$, $K_y^{y^\prime}(\theta)$, $K_z^{z^\prime}(\theta)$, $K_x^{z^\prime}(\theta)$, and $K_z^{x^\prime}(\theta)$. 
In Fig. \ref{fig:spin-trns-prime-19mev}, calculations of these observables at $E_n = 19.0$ MeV for $60^\circ \le \theta \le 140^\circ$, where calculations of different models scatter,  are displayed comparing with experimental data of $K_y^{y^\prime}(\theta)$ for $n$-$d$ scattering \cite{He98} and those of $K_y^{y^\prime}(\theta)$ and $K_z^{x^\prime}(\theta)$ for $p$-$d$ scattering \cite{Sy94}. 

As concerns $K_y^{y^\prime}(\theta)$, the calculations and the $n$-$d$ data show similar tendencies to those of $T_{21}(\theta)$ in Fig. \ref{fig:t21-28mev}, namely the smallness of the SO-3NF effect and the partial cancellation of the 2$\pi$E-3NF effect by the tensor inversion.
It is likely that the tensor inversion effect works nicely in reproducing the $n$-$d$ data. 
However, the position of the $n$-$d$ data point at $\theta=110^\circ$ in Fig. \ref{fig:spin-trns-prime-19mev} (b), which is above the $p$-$d$ data point, is contradictive to Kohn variational principle calculations with or without including a Coulomb force effect \cite{Wi06}. 
Therefore, it is not conclusive whether the tensor inverse effect is favorable or not until more experimental data are accumulated and/or Coulomb effects are fixed. 

Fig. \ref{fig:spin-trns-prime-19mev} displays another features of the observables:
$K_x^{x^\prime}(\theta)$ and $K_z^{z^\prime}(\theta)$ reveal the dependence on both of the tensor inversion and the SO-3NF. 
On the other hand, $K_x^{z^\prime}(\theta)$ and $K_z^{x^\prime}(\theta)$ do a scaling behavior, i.e., the results with models that equally reproduce the triton binding energy almost agree with each other. 

%
Some of these characteristics may be understood in the following way:
In Ref. \cite{Is02}, $N$-$d$ observables $K_x^x(\theta)$, etc. are analyzed in terms of $N$-$d$ scattering amplitudes, which are decomposed into scalar, vector, tensor, etc. in spin space. 
Analyses in Appendix B 8 of Ref. \cite{Is02} show that $K_x^{z}(\theta)$ and $K_z^{x}(\theta)$ are proportional to both of vector and tensor components in $N$-$d$ amplitudes, which are sensitive to spin-orbit and tensor forces, respectively, 
and that $K_x^x(\theta)$ and $K_z^z(\theta)$ are governed by scalar components leading to the scaling behavior because of their sensitivity to overall attraction of the nuclear forces.
Note that we deal with two different sets of the observables referring to two different coordinate systems, $(x,y,z)$ and $(x^\prime,y^\prime,z^\prime)$. 
The former (latter) system is defined so that the $z$-axis ($z^\prime$-axis) is oriented to the direction of the beam (observed) particle. 
In the $N$-$d$ elastic scattering, a scattering angle of $\theta=120^\circ$, around which we are interested in, corresponds to a scattering angle of $90^\circ$ in laboratory system. 
In this particular angle, it is easily shown that relations: 
$\hat{x}^\prime = -\hat{z}$, $\hat{y}^\prime=\hat{y}$, and $\hat{z}^\prime= \hat{x}$ hold, and thus we have,
$K_x^{x^\prime}(\theta) = -K_x^{z}(\theta)$, 
$K_z^{z^\prime}(\theta) = K_z^{x}(\theta)$, 
$K_x^{z^\prime}(\theta) = K_x^{x}(\theta)$, and
$K_z^{x^\prime}(\theta) = -K_z^{z}(\theta)$.
This lead to the features of $K_x^{x^\prime}(\theta)$, $K_z^{z^\prime}(\theta)$, $K_x^{z^\prime}(\theta)$, and $K_z^{x^\prime}(\theta)$ around $\theta=120^\circ$ are originated from as reflections of those of $K_x^{z}(\theta)$, $K_z^{x}(\theta)$, $K_x^x(\theta)$, and $K_z^z(\theta)$, respectively.


\begin{figure}[tbh]
\includegraphics[scale=0.29]{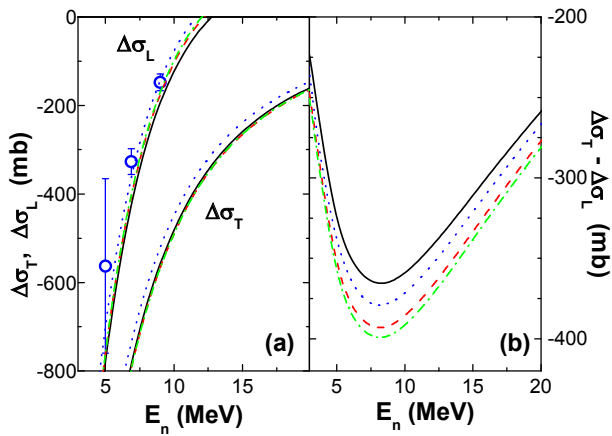}
\caption{(Color online) 
(a) Spin dependent cross section differences, $\Delta\sigma_L$ and $\Delta\sigma_T$, and (b) their difference $\Delta\sigma_T-\Delta\sigma_L$.
Dotted (blue) curves denotes the AV18 calculations; 
dotted-dashed (green) curves AV18+BR$_{680}$; 
dashed (red) curves AV18+BR$_{680}$+SO;
solid (black) curves AV18+BR$_{800}$+C+T+SO.
Experimental data of $\Delta\sigma_L$ are taken from \protect\cite{Fo06}.
\label{fig:del-sig-tl-br}
}
\end{figure}

Final examples of observables that are sensitive to our model 3NFs are spin-dependent total cross section differences in $\vec{n}$-$\vec{d}$ scattering, $\Delta\sigma_L$ and $\Delta\sigma_T$ \cite{Fo06,Is03c}.
These observables are particularly interesting because they are related directly to the imaginary part of $n$-$d$ scattering amplitudes at forward angle by the optical theorem.
In Fig. \ref{fig:del-sig-tl-br} (a),  we show calculations of $\Delta\sigma_L$ and $\Delta\sigma_T$ as well as recent measurements of $\Delta\sigma_L$ \cite{Fo06}. 
In Ref. \cite{Is03c}, it is pointed out that the difference $\Delta\sigma_T-\Delta\sigma_L$ is proportional to the imaginary part of a tensor component in the $n$-$d$ scattering amplitudes at forward angle. 
As Fig. \ref{fig:del-sig-tl-br} (b) shows, the calculations of the difference for AV18+BR$_{800}$+C+T+SO and AV18+BR$_{680}$+SO lie oppositely to the AV18 calculation, which means the tensor inversion effect is observed clearly.
Thus, precise measurements of these observables are quite interesting to obtain information about tensor components in 3NFs.


In summary, we have studied effects of the spin-dependence in nuclear interactions on $N$-$d$ polarization observables using a model 3NF to be added to the Argonne V$_{18}$ 2NF and the Brazil 2$\pi$E-3NF. 
The model 3NF consists of spin-independent, spin-orbit, and tensor components, which are essential in reproducing the 3N binding energy, the proton vector analyzing power $A_y(\theta)$ and the deuteron tensor analyzing power $T_{21}(\theta)$ in $p$-$d$ scattering at 3 MeV, respectively. 
Effects of the model 3NF on some $N$-$d$ polarization observables at higher energies are examined.
The spin-orbit component in the model 3NF plays a significant role in reproducing $A_y(\theta)$. 
The tensor component has nonnegligible effects in $T_{21}(\theta\approx90^\circ)$, $A_y(\theta\approx90^\circ)$, and $K_y^{y^\prime}(\theta\approx120^\circ)$, although we need more experimental data to confirm whether its effect is favorable or not. 
$K_x^{x^\prime}(\theta\approx120^\circ)$ and $K_z^{z^\prime}(\theta\approx120^\circ)$ depend on both of the spin-orbit and the tensor components. 
Effects of the tensor component are clearly seen in the spin-dependent $n$-$d$ cross section differences. 
Thus, further experimental studies of these observables are expected to improve our knowledge for the spin-dependence of three-nucleon forces.

If the model 3NF presented in this work were really successful in explaining all of experimental data, it is interesting to see what kind of realistic 3NFs are simulated by this.
In Ref. \cite{Is06}, 3NFs arising from the exchange of a pion and a scalar object, such as $\sigma$, $\omega$, or a scalar part of two-pion exchange, are examined. 
It is shown that these type of 3NFs give similar effects to the tensor inversion qualitatively, but not sufficient quantitatively. 
Of course, this is not conclusive due to a large amount of uncertainties in the description of such processes, and further study is required.

\begin{acknowledgments}
The numerical calculations were supported by Research Center for Computing and Multimedia Studies, Hosei University, under Project No. lab0003.
\end{acknowledgments}


\end{document}